\documentclass[aps,prd,preprint,amsmath,amssymb,
nofootinbib,eqsecnum,showpacs,preprint,tightenlines]{revtex4}
\usepackage{epsfig}
\usepackage{rotating}
\usepackage{bm}
\newcommand{\be}{\begin{equation}}
\newcommand{\ee}{\end{equation}}
\newcommand{\bea}{\begin{eqnarray}}
\newcommand{\eea}{\end{eqnarray}}
\begin{document}
\title{Local Casimir Energies for a Thin Spherical Shell}
\author{In\'es Cavero-Pel\'aez}
\email{cavero@nhn.ou.edu}

\author{Kimball A. Milton}\thanks{On sabbatical at: Department of
Physics, Washington University, St. Louis, MO 63130-4899}
\email{milton@nhn.ou.edu}
\homepage{www.nhn.ou.edu/

\author{Jeffrey Wagner}
\email{wagner@nhn.ou.edu}
\affiliation{Department of Physics and Astronomy, University of Oklahoma,
Norman, OK 73019-2061}

\date{\today}

\pacs{03.70.+k, 11.10.Gh, 03.65.Sq}

\begin{abstract}
The local Casimir energy density for a massless scalar field
associated with step-function potentials in a $3+1$
dimensional spherical geometry is considered. The potential is chosen
to be zero except in a shell of thickness $\delta$, where it has
height $h$, with the constraint
$h\delta=1$.  In the limit of zero thickness, an ideal
$\delta$-function shell is recovered.  In this limit,
the behavior of the energy density
as the surface of the shell is approached is studied in both the strong
and weak coupling regimes.  The former case corresponds to the well-known
Dirichlet shell limit.  New results, which shed light on the nature
of surface divergences and on the energy contained within the shell,
are obtained in the weak coupling limit,
and for a shell of finite thickness. In the case of zero
thickness, the energy has a
contribution not only from the local energy density, but from an energy
term residing entirely on the surface.  It is shown that the latter coincides
with the integrated local energy density within the shell.
We also study the dependence of local and global 
quantities on the conformal parameter.  In particular new insight is
provided on the reason for the divergence in the global Casimir energy
in third order in the coupling.

\end{abstract}

\maketitle
\section{Introduction}
The subject of local energy density associated with the confinement of
quantum fields by surfaces has a rather long history.  For example,
Brown and Maclay \cite{Brown:1969na} computed the vacuum expectation
value of the electromagnetic
energy-momentum tensor between two parallel perfectly conducting plates, which 
is twice that of a conformally coupled scalar field satisfying Dirichlet or
Neumann boundary conditions on the plates separated by a distance $a$ in
the $x$ direction, namely
\be
\langle T^{\mu\nu}\rangle=\frac{\pi^2}{1440 a^4}\mbox{diag}\,(-1,-3,1,1),
\ee
which corresponds precisely to the attractive energy or pressure found
by Casimir in the same situation \cite{Casimir:1948dh}.  If a nonconformal
scalar stress tensor is used, a position-dependent term in the stress tensor
appears, which does not contribute to either the total energy or the pressure
on the plates \cite{Milton:2002vm,Actor:1996zj}.

Local surface divergences were first discussed for arbitrary smooth boundaries
by Deutsch and Candelas \cite{Deutsch:1978sc}.  They found cubic divergences
in the energy density as one approaches the surface; for example, outside
a Dirichlet sphere (that is, for a conformally-coupled
scalar field satisfying Dirichlet boundary
conditions on the surface) the energy density diverges as
\be
r\to a+:\quad u\sim \frac1{360\pi^2}\frac1{a(r-a)^3},\label{dc1}
\ee
where $a$ is the radius of the sphere, and $r$ is the distance from the
center of the sphere.
This raises the question: How can it be that the total Casimir energy of
a Dirichlet sphere (or a perfectly conducting sphere in electrodynamics)
is finite?  The electromagnetic case is the well-known one first calculated
by Boyer \cite{Boyer:1968uf}
\be
\mathcal{E}^{EM}=\frac{0.04618}a,\label{boyer}
\ee
while the scalar case was first
worked out by Bender and Milton \cite{Bender:1994zr},
\be
\mathcal{E}^S=\frac{0.002817}a.\label{bender}
\ee
Thus there has been a feeling since the time of Deutsch and Candelas 
that there was something suspect about the calculations of Casimir self
energies of ideal closed boundaries.  (We note that there is now a proof
that any such smooth perfectly conducting boundary possesses a finite 
electromagnetic Casimir energy \cite{graf}.
Whether such an idealized limit is physical is, of course, another question.) 
This suspicion has been
recently intensified by a series of talks and papers by Graham et 
al.~\cite{Graham:2003ib}.  The essential outcome of their analysis is that
for a $\delta$-function sphere, described by the following Lagrangian
for a massless scalar field,
\be
\mathcal{L}=-\frac12\partial_\mu\phi\partial^\mu\phi-\frac12
\frac{\lambda}{a^2}\delta(r-a)\phi^2,\label{deltapot}
\ee
a divergence occurs in third order in $\lambda$.  (They claimed a divergence
in second order, but that was spurious 
\cite{Milton:2002vm,Milton:2004ya,Milton:2004vy}.)  This
divergence in fact was discovered much earlier by Bordag, Kirsten, and
Vassilevich \cite{Bordag:1998vs}, and possible ways of dealing with it
have been suggested \cite{Bordag:2004rx,Scandurra:1998xa}.
(Recently, the effects of the spherical $\delta$-function potential
at finite temperature have been considered by McKenzie-Smith and Naylor
\cite{Mckenzie-Smith:2004hg}.)
Objections complementary to those of Ref.~\cite{Graham:2003ib} have also
been voiced by Barton \cite{barton}, all of which raise doubts concerning the
physical relevance of results such as those in Eqs.~(\ref{boyer}) and
(\ref{bender}).

Clearly, then, there are issues here unresolved.  The purpose of the
present paper is to examine the local energy density for a scalar field
in the presence of a spherically symmetric background, but as suggested
by Ref.~\cite{Graham:2003ib} not so singular as a delta function.  So we
consider a spherical shell, with a finite step potential.  In the limit
as the height goes to infinity and the width to zero we recover the 
$\delta$-function sphere.  This enables us to explore how the quantum vacuum
energy of the shell itself contributes to the total energy of the 
configuration.  In the next section, we will construct the Green's
function for such a sphere, and in the following section the local energy
density (for an arbitrarily coupled scalar) inside, outside, and within the
region of the shell.  In Sec.~\ref{Sec:surfdiv} we will consider the
thin shell limit and examine how the energy density diverges as the surface is
approached. As expected, the divergence in the energy density as the surface
is approached is weakest for the conformally coupled case.
We also study how the planar divergences within and outside a thick shell
pass over to the divergences exterior to a thin shell.
 In Sec.~\ref{Sec:toten} we will compute the energy of each region,
and thereby determine the total energy of the spherical geometry.  There
we will see that the former is constructed not only from
 the local energy density,
which diverges as the surface is approached, but also contains an energy lying
entirely on the surfaces bounding the region,
except for a special, nonconformal value of the conformal
parameter.  For finite thickness of the shell, the surface energies cancel
across each surface in constructing
the total energy.  However, the thin shell limit
corresponds to a singular potential, for which there is a
net effective surface energy for the sum of the interior and exterior
regions of the shell which is identified as the integrated energy
density within the shell.

\section{Green's function for $\lambda$ sphere}
As discussed in Refs.~\cite{Milton:2004ya,Milton:2004vy,Graham:2002yr} 
for parallel planes,  we consider the potential
\begin{subequations}\label{smoothpot}
\be \mathcal{L}_{\rm int}=-\frac\lambda{2a^2}\phi^2\sigma(r),
\ee
where
\be
\sigma(r)=\left\{\begin{array}{cc}
0,&r<a_-,\\
h,&a_-<r<a_+,\\
0,&a_+<r.\end{array}\right.
\ee
\end{subequations}
Here $a_\pm=a\pm\delta/2$, and we set $h\delta=1$.  
Compared to Refs.~\cite{Milton:2004ya,Milton:2004vy}, we have
changed the dimensions of $\lambda$ so that the total energy of
interaction does not explicitly refer to the radius $a$. In the limit
as $\delta\to0$ (or $h\to\infty$) we recover the $\delta$-function sphere
considered first in Ref.~\cite{Bordag:1998vs}.

A straightforward solution of the Green's function equation, 
for a massless particle, with $\kappa^2=-\omega^2$,
\be
\left(-\nabla^2+\kappa^2+\frac\lambda{a^2}
\sigma\right)\mathcal{G}(\mathbf{r,r'})=\delta(\mathbf{r-r'}),
\ee
is given in terms of the reduced Green's function,
\be
\mathcal{G}(\mathbf{r,r'})=
\sum_{lm}g_l(r,r')Y_{lm}(\theta,\phi)Y_{lm}^*(\theta',\phi').
\label{redgf}
\ee
The latter is expressed in terms of the modified Ricatti-Bessel functions, 
\be
s_l(x)=\sqrt{\frac{\pi x}2}I_{l+1/2}(x),\quad
e_l(x)=\sqrt{\frac{2x}\pi}K_{l+1/2}(x),
\ee
as follows, outside of the shell,
\begin{subequations}\label{goutside}
\bea
r,r'<a_-:\quad g_l&=&\frac1{\kappa rr'}\left[s_l(\kappa r_<)e_l(\kappa r_>)
-\frac{\tilde\Xi}{\Xi}s_l(\kappa r)s_l(\kappa r')\right],\label{gl1}
\\
r,r'>a_+:\quad g_l&=&\frac1{\kappa rr'}\left[s_l(\kappa r_<)e_l(\kappa r_>)
-\frac{\hat\Xi}{\Xi}e_l(\kappa r)e_l(\kappa r')\right],\label{gl2}
\eea
\end{subequations}
where the denominator is
\bea
\Xi&=&[\kappa s_l'(\kappa a_-)e_l(\kappa' a_-)-\kappa' s_l(\kappa a_-)
e_l'(\kappa'a_-)][\kappa' e_l(\kappa a_+)s_l'(\kappa'a_+)
-\kappa e_l'(\kappa a_+)s_l(\kappa'a_+)]\nonumber\\
&&\mbox{}-[\kappa s_l'(\kappa a_-) s_l(\kappa' a_-)-\kappa's_l(\kappa a_-)
s_l'(\kappa'a_-)][\kappa' e_l(\kappa a_+)e_l'(\kappa'a_+)-\kappa
e_l'(\kappa a_+)e_l(\kappa' a_+)].\nonumber\\
\label{xi}
\eea
The numerator in (\ref{gl1}),
$\tilde\Xi$, is obtained from $\Xi$ by replacing $s_l(\kappa a_-)\to
e_l(\kappa a_-)$:
\bea
\tilde\Xi&=&[
\kappa {e_l'(\kappa a_-)}
e_l(\kappa' a_-)-{\kappa' e_l(\kappa a_-)}
e_l'(\kappa'a_-)]
[\kappa' e_l(\kappa a_+)s_l'(\kappa'a_+)
-\kappa e_l'(\kappa a_+)s_l(\kappa'a_+)]\nonumber\\
&&\mbox{}-[\kappa{e_l'(\kappa a_-)} 
s_l(\kappa' a_-)-\kappa'{e_l(\kappa a_-)}
s_l'(\kappa'a_-)][\kappa' e_l(\kappa a_+)e_l'(\kappa'a_+)-\kappa
e_l'(\kappa a_+)e_l(\kappa' a_+)],\nonumber\\
\eea
while that in (\ref{gl2}), $\hat\Xi$, is obtained from $\Xi$ by replacing
$e_l(\kappa a_+)\to s_l(\kappa a_+)$:
\bea
\hat\Xi&=&[
\kappa s_l'(\kappa a_-)
e_l(\kappa' a_-)-\kappa' s_l(\kappa a_-)
e_l'(\kappa'a_-)][\kappa' {s_l(\kappa a_+)}
s_l'(\kappa'a_+)
-\kappa{s_l'(\kappa a_+)}
s_l(\kappa'a_+)]\nonumber\\
&&\mbox{}-[\kappa s_l'(\kappa a_-)s_l(\kappa' a_-)
-\kappa's_l(\kappa a_-)
s_l'(\kappa'a_-)][\kappa'{s_l(\kappa a_+)}
e_l'(\kappa'a_+)-\kappa
{s_l'(\kappa a_+)}e_l(\kappa' a_+)].\nonumber\\
\eea
Here $\kappa'=\sqrt{\kappa^2+\lambda h/a^2}$.  It might be noticed
that under the interchange
\bea
s_l(\kappa a_\mp)\leftrightarrow e_l(\kappa a_\pm),
\eea and similarly for functions depending on $\kappa'$, $\Xi$ is unchanged,
while $\hat\Xi\leftrightarrow\tilde \Xi$.

The Green's function within the shell has a somewhat more complicated form.
For $a-\frac\delta2<r,r'<a+\frac\delta2$ we find
\bea
g_l&=&\frac1{\kappa' rr'}\bigg\{s_l(\kappa'r_<)e_l(\kappa'r_>)-\frac1\Xi\bigg[
A[s_l(\kappa'r)e_l(\kappa'r')+s_l(\kappa'r')e_l(\kappa'r)]\nonumber\\
&&\mbox{}+Bs_l(\kappa'r')s_l(\kappa'r)
+Ce_l(\kappa'r')e_l(\kappa'r)\bigg]\bigg\},\label{2.10}
\eea
where
\begin{subequations}\label{2.10a}
\bea
A&=&[\kappa e_l'(\kappa a_+)e_l(\kappa'a_+)-\kappa'e_l(\kappa a_+)
e_l'(\kappa'a_+)][\kappa s_l'(\kappa a_-)s_l(\kappa'a_-)
-\kappa's_l(\kappa a_-)s_l'(\kappa'a_-)],\nonumber\\
\label{2.10aa}\\
B&=&-[\kappa e_l'(\kappa a_+)e_l(\kappa'a_+)
-\kappa'e_l(\kappa a_+)e'_l(\kappa' a_+)]
[\kappa s_l'(\kappa a_-)e_l(\kappa' a_-)-\kappa' s_l(\kappa a_-)
e_l'(\kappa' a_-)],\nonumber\\
\label{2.10ab}\\
C&=&-[\kappa e_l'(\kappa a_+)s_l(\kappa'a_+)
-\kappa'e_l(\kappa a_+)s'_l(\kappa' a_+)]
[\kappa s_l'(\kappa a_-)s_l(\kappa' a_-)-\kappa' s_l(\kappa a_-)
s_l'(\kappa' a_-)].\nonumber\\
\label{2.10ac}
\eea
\end{subequations}
A check of this occurs when we take the interior radius to zero, $a_-\to0$,
for then we recover the known result for (the TE part of)
a solid dielectric ball \cite{Milton:1979yx}.  
Then $s_l(\kappa a_-)\to0$, and the $A$ and $C$
coefficients above vanish.  The denominator also simplifies, and we obtain
the following form for the Green's function within and without the ball,
of radius $a_+$ now called $a$:
\begin{subequations}
\bea
g_l(r,r')&=&\frac1{\kappa' rr'}[s_l(\kappa'r_<)e_l(\kappa'r_>)-\tilde\Omega
s_l(\kappa'r)s_l(\kappa'r')],\quad r,r'<a,\\
&=&\frac1{\kappa rr'}[s_l(\kappa r_<)e_l(\kappa r_>)-\hat\Omega
e_l(\kappa r)e_l(\kappa r')],\quad r,r'>a,
\eea
\end{subequations}
where
\begin{subequations}
\bea
\tilde\Omega&=&\lim_{a_-\to0}\frac{B}{\Xi}=\frac{\kappa'e_l(\kappa a)
e_l'(\kappa'a)-\kappa e_l'(\kappa a)e_l(\kappa'a)}{\kappa'e_l(\kappa a)
s_l'(\kappa'a)-\kappa e_l'(\kappa a)s_l(\kappa'a)},\\
\hat\Omega&=&\lim_{a_-\to0}\frac{\hat\Xi}{\Xi}=\frac{\kappa's_l(\kappa a)
s_l'(\kappa'a)-\kappa s_l'(\kappa a)s_l(\kappa'a)}{\kappa'e_l(\kappa a)
s_l'(\kappa'a)-\kappa e_l'(\kappa a)s_l(\kappa'a)}.
\eea
\end{subequations}
The required symmetry between the inside and outside forms of the energy,
obtained by interchanging $s_l\leftrightarrow e_l$ and $\kappa\leftrightarrow
\kappa'$, is now manifest.  This result for the Green's function is exactly
that found in Ref.~\cite{Milton:1979yx} for the TE (scalar) part of the
dielectric Green's function with $\mu=1$, 
$\kappa'=\sqrt{\varepsilon'/\varepsilon}\kappa$.

We will now use the Green's function to compute the energy density.

\section{Energy Density}
We can calculate the local energy density from the stress tensor:
\be
T^{\mu\nu}=\partial^\mu\phi\partial^\nu\phi-g^{\mu\nu}\mathcal{L}
-\xi(\partial^\mu\partial^\nu-g^{\mu\nu}\partial^2)\phi^2,\label{stresstensor}
\ee
from which the energy density follows:
\be
T^{00}=\frac12\left[\partial^0\phi\partial^0\phi+\bm{\nabla}\phi\cdot\bm{\nabla}\phi
+\frac{\lambda}{a^2}\sigma\phi^2\right]-\xi\nabla^2\phi^2,
\ee
where the conformal value is given by $\xi=1/6$. 
To obtain the vacuum expectation values at one loop, we use the 
identification with the classical Green's function
\be
\langle\phi(x)\phi(x')\rangle=\frac1i G(x,x').
\ee
The energy density thus is, within or outside the shell,
\be
\langle T^{00}\rangle=\frac1{2i}\left(\partial^0\partial^{\prime0}+\bm{\nabla}
\cdot\bm{\nabla}'+\left\{\begin{array}{c}\lambda h/a^2\\0\end{array}\right\}\right)
G(x,x')\bigg|_{x'=x}-\frac\xi{i}\nabla^2G(x,x).
\ee

When we insert the partial wave decomposition of the Green's function
(\ref{redgf}), we
encounter, in terms of the angle $\gamma$ between the two directions
$(\theta,\varphi)$, $(\theta',\varphi')$,
\be
\cos\gamma=\cos\theta\cos\theta'+\sin\theta\sin\theta'\cos(\varphi-\varphi'),
\ee
the evaluation
\be
\bm{\nabla}\cdot\bm{\nabla}'P_l(\cos\gamma)\bigg|_{\theta=\theta',\phi=\phi'}
=\frac1{r^2}2P_l'(1)=\frac{l(l+1)}{r^2}.
\ee
Therefore, the expression for the energy density is immediately reduced
to (inside or outside the shell, but not within it)
\be
\langle T^{00}\rangle =\int_0^\infty \frac{d\kappa}{2\pi}\sum_{l=0}^\infty
\frac{2l+1}{4\pi}\left\{\left[-\kappa^2+\partial_r\partial_{r'}+
\frac{l(l+1)}{r^2}
\right]g_{l}(r,r')\bigg|_{r'=r}-2\xi\frac1{r^2}\frac{\partial}{\partial r}
r^2\frac{\partial}{\partial r}g_l(r,r)\right\}.\label{t00}
\ee
Note that there has been no need of some dubious argument
(such as appears in Ref.~\cite{Milton:2004vy}, Eq.~(4.21)) concerning
partial integration in the angular coordinates.

We insert the Green's function in the exterior region, but delete the
free part, 
\bea
 g^0_l&=&\frac1{\kappa rr'}s_l(\kappa r_<)e_l(\kappa r_>),
\eea
which corresponds to the {\em bulk energy\/} which would be
present if either medium filled all of space, leaving us with for $r>a_+$
\bea
u(r)&=&-(1-4\xi)\int_0^\infty \frac{d\kappa}{2\pi}\sum_{l=0}^\infty 
\frac{2l+1}{4\pi}\frac{\hat\Xi}{\Xi}\bigg\{\frac{e_l^2(\kappa r)}{\kappa r^2}
\left[-\kappa^2\frac{1+4\xi}{1-4\xi}+\frac{l(l+1)}{r^2}+\frac1{r^2}
\right]\nonumber\\
&&\quad\mbox{}-\frac2{r^3}e_l(\kappa r)e_l'(\kappa r)+\frac\kappa
{r^2}e_l^{\prime2}(\kappa r)\bigg\}.\label{uout}
\eea
Inside the shell, $r<a_-$, the energy is given by a similar expression
obtained from Eq.~(\ref{uout}) by replacing $e_l$ by $s_l$ and $\hat\Xi$
by $\tilde \Xi$. 

Within the shell, $a_-<r<a_+$, the energy density is given by a somewhat
more complicated formula:
\bea
u(r)&=&-(1-4\xi)\int_0^\infty \frac{d\kappa}{2\pi}\sum_{l=0}^\infty
\frac{2l+1}{4\pi}\frac1{\kappa'r^2}
\frac{1}{\Xi}\bigg\{
\left[-\kappa^{\prime 2}\frac{1+4\xi}{1-4\xi}+\frac{l(l+1)+1}{r^2}+
\frac{2\lambda h}{a^2}\frac{1}{1-4\xi}
\right]\nonumber\\
&&\quad\times[2A s_l(\kappa' r)e_l(\kappa'r)+Bs_l^2(\kappa'r)+Ce_l^2(\kappa'r)]
\nonumber\\
&&\quad\mbox{}-\frac{2\kappa'}{r}[A(s_l(\kappa' r)e_l'(\kappa' r)+
e_l(\kappa'r)s_l'(\kappa'r))+Bs_l(\kappa'r)s_l'(\kappa'r)+Ce_l(\kappa'r)
e_l'(\kappa'r)]\nonumber\\
&&\quad\mbox{}+\kappa^{\prime2}[2As_l'(\kappa'r)e_l'(\kappa'r)+B s_l^{\prime2}
(\kappa'r)+Ce_l^{\prime2}(\kappa' r)]\bigg\}.\label{ushell}
\eea
Here $A$, $B$, and $C$ are the coefficients given in
Eq.~(\ref{2.10a}).

\section{Surface Divergences in the Energy Density}\label{Sec:surfdiv}
\subsection{Exterior divergences for $\delta=0$}\label{sec:ext}
We want to examine the singularity structure as $r\to a_+$
from the outside.   For this purpose
we use the leading uniform asymptotic expansion, $l\to\infty$,
\bea
 e_l(x)&\sim& \sqrt{zt}\,e^{-\nu \eta},\quad
s_l(x)\sim\frac12\sqrt{zt}\,e^{\nu \eta},\nonumber\\
e_l'(x)&\sim&-\frac1{\sqrt{zt}}\,e^{-\nu\eta},\quad s'_l(x)\sim
\frac12\frac1{\sqrt{zt}}\,e^{\nu \eta},\label{uae}
\eea
where ($\nu=l+1/2$)
\bea x=\nu z,\quad t=(1+z^2)^{-1/2},\quad \frac{d\eta}{dz}=\frac1{zt}.
\eea
Let us first consider the thin shell limit, $\delta\to 0$, $h\delta=1$, where
it is easy to check that
\be
\frac{\hat\Xi}{\Xi}\to\frac{\frac\lambda{\kappa a^2} s^2_l(\kappa a)}
{1+\frac\lambda{\kappa a^2}e_l(\kappa a)s_l(\kappa a)},
\ee
which is exactly the coefficient occurring in the $\delta$-function 
potential (\ref{deltapot}).  
There are two simple limits of this, strong and weak coupling:
\begin{subequations}
\bea
\frac{\lambda}a \to\infty:\quad\frac{\hat\Xi}{\Xi}&\to&
\frac{s_l(\kappa a)}{e_l(\kappa a)},\\
\frac{\lambda}a\to0:\quad  \frac{\hat\Xi}{\Xi}&\to&
\frac\lambda{\kappa a^2}s_l^2(\kappa a),
\eea
\end{subequations}
if we assume that the relevant scale of $\kappa$ is $1/a$, since we
expect that the significant values of $\kappa$ are determined by the
argument of the Bessel functions in Eq.~(\ref{uout}).

In either case, we carry out the asymptotic sum over angular momentum using
Eq.~(\ref{uae}) and
\bea
\sum_{l=0}^\infty e^{-\nu\chi}=\frac1{2\sinh\frac\chi2}.\label{sumoverl}
\eea
Here ($r\approx a$)
\be
\chi=2\left[\eta(z)-\eta\left(z\frac{a}r\right)\right]\approx
2z\frac{d\eta}{dz}(z)\frac{r-a}r=\frac2t\frac{r-a}r.
\ee
The remaining integrals over $z$ are elementary, and in this 
way we find that the leading divergences
in Eq.~(\ref{uout}) are as $r\to a+$,
\begin{subequations}
\bea
\frac\lambda{a}\to\infty:\quad u&\sim&
-\frac1{16\pi^2}\frac{1-6\xi}{(r-a)^4},\label{dsphere1}\\
\frac\lambda{a}\to0:\quad  u^{(n)}&\sim&
\left(-\frac\lambda{a}\right)^n\frac{\Gamma(4-n)}{96\pi^2a^4}(1-6\xi)
\left(\frac{a}{r-a}\right)^{4-n},\quad n<4,\label{surfdivn}
\eea
\end{subequations} 
where the latter is  the leading divergence in order $n$. 
These results clearly seem to demonstrate the virtue of the conformal 
value of $\xi=1/6$; but see below.
(The  value for the Dirichlet sphere (\ref{dsphere1}) first appeared
in Ref.~\cite{Deutsch:1978sc}; it
recently was rederived in Ref.~\cite{Schwartz-Perlov:2005ds},
where, however, the subdominant term, the leading
term if $\xi=1/6$, namely (\ref{sdsc}), was not calculated.
Of course, this result is the same as the surface divergence
encountered for parallel Dirichlet plates \cite{Milton:2002vm}.)

Thus, for $\xi=1/6$ we must keep subleading terms. 
This includes keeping the subdominant term in $\chi$,
\be
\chi\approx\frac2t\frac{r-a}r-t\left(\frac{r-a}r\right)^2,
\ee
the distinction between $t(z)$ and $\tilde t=t(\tilde z=za/r)$,
\be
\tilde z\tilde t\approx zt-t^3z\frac{r-a}r,
\ee
as well as the next term in the uniform asymptotic expansion of the
Bessel functions,
\begin{subequations}
\bea
s_l(x)&\sim&\frac12\sqrt{zt}\,e^{\nu\eta}\left(1+\frac1\nu u_1(t)+
O(\nu^{-2})\right),\\
e_l(x)&\sim&\sqrt{zt}\,e^{-\nu\eta}\left(1-\frac1\nu u_1(t)+
O(\nu^{-2})\right),\\
s'_l(x)&\sim&\frac12\frac1{\sqrt{zt}}\,e^{\nu\eta}\left(1+\frac1\nu
v_1(t)+O(\nu^{-2})\right),\\
e'_l(x)&\sim&-\frac1{\sqrt{zt}}\,e^{-\nu\eta}\left(1-\frac1\nu v_1(t)
+O(\nu^{-2})\right),
\eea
\end{subequations}
where
\be
u_1(t)=\frac{3t-5t^3}{24},\quad v_1(t)=\frac{3t+7t^3}{24}.
\ee
Including all this, it is straightforward to recover the
well-known result (\ref{dc1}) \cite{Deutsch:1978sc}
for strong coupling (Dirichlet boundary conditions):
\be
 \frac\lambda{a}\to\infty:\quad  u\sim
\frac1{360\pi^2}\frac1{a(r-a)^3},\label{sdsc}
\ee
Following the same process for
weak coupling, we find that the 
 leading divergence in order $n$, $1\le n<3$, is ($r\to  a\pm$)
\be
\lambda\to 0:\quad u^{(n)}\sim\left(\frac\lambda{a^2}\right)^{n}
\frac1{1440\pi^2}\frac1{a(a-r)^{3-n}}(n-1)(n+2)\Gamma(3-n).\label{sdwc}
\ee  
Note that the subleading $O(\lambda)$ term again vanishes.
Both Eqs.~(\ref{sdsc}) and (\ref{sdwc}) apply 
for the conformal value $\xi=1/6$.

\subsection{Divergences within and outside the shell for $\delta\ne0$}
\label{sec:int}
Now we take finite $\delta$, and examine the energy density (\ref{ushell})
within the shell. Again, the singularities in the energy density
as $r$ approaches $a_+$ from within the shell are
revealed by examining the behavior as $\kappa$, $\kappa'\to\infty$.  
From the leading uniform asymptotic expansion (\ref{uae}), the denominator
(\ref{xi}) becomes ($l\to\infty$)
\bea
\Xi&\sim&e^{\nu[\eta_--\eta'_-+\eta'_+-\eta_+]}\frac14\left(\kappa
\sqrt{\frac{z_-'t_-'}{z_-t_-}}+\kappa'\sqrt{\frac{z_-t_-}{z_-'t_-'}}
\right)
\left(\kappa'\sqrt{\frac{z_+t_+}{z_+'t_+'}}+\kappa\sqrt{\frac{z_+'t_+'}
{z_+t_+}}\right)\nonumber\\
&&\mbox{}+e^{\nu[\eta_-+\eta_-'-\eta_+'-\eta_+]}\frac14\left(\kappa
\sqrt{\frac{z_-'t_-'}{z_-t_-}}-\kappa'\sqrt{\frac{z_-t_-}{z_-'t_-'}}\right)
\left(\kappa'\sqrt{\frac{z_+t_+}{z_+'t_+'}}-\kappa\sqrt{\frac{z_+'t_+'}
{z_+t_+}}\right),\label{den}
\eea
where $z_-=\kappa a_-/\nu$, $z_+'=\kappa' a_+/\nu$, 
$\eta'_+=\eta(z'_+)$, etc.  Because $\eta'_+>\eta'_-$,
it is clear that the first term in Eq.~(\ref{den}) dominates, so as 
$l\to \infty$,
\be
\Xi\sim e^{\nu[\eta_--\eta'_-+\eta'_+-\eta_+]}\frac14\kappa^2\sqrt{
\frac{z_-'t_-'z_+'t_+'}{z_-t_-z_+t_+}}\left(1+\frac{t_-}{t_-'}\right)
\left(1+\frac{t_+}{t_+'}\right).
\ee
As for the
numerator in Eq.~(\ref{ushell}) it is clear from Eq.~(\ref{uae}) 
that the terms proportional to
$s_l^2(\kappa'r)$ and $s_l^{\prime2}(\kappa'r)$ dominate, and so again we need
only examine the $B$ coefficient (\ref{2.10ab}), which is approximated by
\be
B\sim e^{\nu[\eta_--\eta'_--\eta'_+-\eta_+]}\frac12\kappa^2\sqrt{
\frac{z_-'t_-'z_+'t_+'}{z_-t_-z_+t_+}}\left(1+\frac{t_-}{t_-'}\right)
\left(1-\frac{t_+}{t_+'}\right).
\ee
Thus the leading coefficient in the Green's function or the energy is
\be
\frac{B}\Xi\sim2 e^{-2\nu\eta_+'}\frac{1-\frac{t_+}{t_+'}}{1+\frac{t_+}{t_+'}}.
\ee

Now we must relate $t_+'$ and $t_+$.  $t_+=(1+z_+^2)^{-1/2}$, and
\bea
t_+'&=&\frac1{\sqrt{1+\frac{\kappa^{\prime2}a_+^2}{\nu^2}}}
=\frac1{\sqrt{1+\frac{a_+^2}{\nu^2}\left(\kappa^2+\frac{\lambda h}{a^2}
\right)}}\nonumber\\
&=&\frac1{\sqrt{(1+z_+^2)\left(1+\frac{\lambda h a_+^2}{\nu^2 a^2(1+z_+^2)}
\right)}}\approx t_+\left(1-\frac{\lambda h}{2\nu^2}\frac{a_+^2}{a^2}t_+^2
\right),
\eea
again as $\nu\to\infty$.  Thus, in this approximation the energy density
(\ref{ushell}) in the shell is 
\be
u\sim \frac{\lambda h}{64\pi^2}\frac{a_+^3}{a^2r^3}\frac1{(a_+-r)^2}
\int_0^\infty d z_+\frac{t_+^4}{t_r}(t_r^{2}-4\xi),
\ee
where we have used the evaluation (\ref{sumoverl}), or
\be
\sum_{l=0}^\infty \nu e^{-\nu\chi}\sim\left(\frac{t_+}2\frac{a_+}{r-a_+}
\right)^2.
\ee
Then in the limit as $r$ approaches the outer surface
\be
u=\langle T^{00}\rangle\sim \frac{\lambda h}{96\pi^2 a^2}\frac1{(a_+-r)^2}
(1-6\xi),\quad r\to a_+-.\label{paradiv1}
\ee
Exactly the same result (with $a_+$ replaced by $a_-$) is obtained as $r$
approaches the inner surface of the shell from within the shell, $r\to a_-+$.

That Eq.~(\ref{paradiv1}) is exactly the expected result is seen by
recalling from 
Ref.~\cite{Milton:2004ya,Milton:2004vy,Graham:2002yr} 
the divergence in the local energy density encountered as one
approaches the surface of a slab, centered on the origin,
 of height $h$ and thickness $\delta$ with
two transverse dimensions from the inside, Eq.~(2.52) of 
Ref.~\cite{Milton:2004ya} (or Eq.~(91) of Ref.~\cite{Graham:2002yr}): 
($\lambda h/a^2\to \lambda h$ there)
\be
\langle T^{00}\rangle\sim\frac{(1-6\xi)\lambda h}{96\pi^2(x-\delta/2)^2},
\quad |x|\to\delta/2.\label{paradiv}
\ee
This was  obtained, as here,
 by looking at the most divergent part of the $\kappa$
integral, and using $\kappa^2\gg \lambda h$.  On physical grounds, we must
encounter the same
divergence here, because if we are very close to the surface locally it
looks flat.  

We obtain the opposite behavior outside the outer surface, where the same 
asymptotic analysis implies
\be
\frac{\hat\Xi}\Xi\sim-\frac12 e^{2\nu \eta_+}\frac{1-\frac{t_+}{t_+'}}
{1+\frac{t_+}{t_+'}}
\sim\frac12e^{2\nu\eta_+}\frac{\lambda h}{2\nu^2}\frac{a_+^2t_+^2}{2a^2}.
\ee
This implies exactly the negative of the divergence found in 
Eq.~(\ref{paradiv1}) as $r$ approaches the surface from the outside.
(Again, the change of sign was found outside the
slab \cite{Milton:2004ya,Graham:2002yr}.)

This, obviously, is quite
different from the surface divergence we encounter in 
Eq.~(\ref{surfdivn})---it has a different
dependence on the distance from the surface, and furthermore it
depends explicitly on $h$.  
Unlike the curvature divergences found in the previous subsection,
the divergences found here are independent of curvature [$\lambda h/a^2$ 
is simply the coupling strength of the potential in Eq.~(\ref{smoothpot})],
it cancels between inside and outside the boundary, and it
vanishes if the conformal coupling is used.  As in the parallel plate
situation, it is without physical consequence therefore, and we will
disregard it in the following, where we wish to concentrate on the
thin shell, $\delta\to0$, limit.

It is, however, easy to provide a heuristic  connection between the divergence
structure seen in Eq.~(\ref{paradiv1}) and that obtained outside a
thin shell, Eq.~(\ref{surfdivn}) for $n=1$. The latter is obtained
from the former by considering the sum of the energy densities
due to the inner and outer surfaces,
$a_\pm\gg r-a_\pm\gg \delta$,
\bea
u(r\to a_++)+u(r\to a_-+)&=&\frac{\lambda h}{96\pi^2a^2}(1-6\xi)\left[
\frac1{(r-a_-)^2}+\frac1{(r-a_+)^2}\right]\nonumber\\
&\to&-\frac{\lambda(1-6\xi)}{48\pi^2a^2(r-a)^3},\quad a_+-a_-=\delta\to0,
\eea
again using $h\delta=1$.  This is exactly the expected divergence, which
therefore has nothing to do with curvature, since the same result may
be obtained either by this argument, or direct calculation, for a thin
($\delta$-function) plane slab.

\section{Total Energy for a Thin Spherical Shell}
\label{Sec:toten}
\subsection{Cancellation of divergences}
The results (\ref{sdwc})
 from Sec.~\ref{sec:ext} for the conformally coupled scalar
for a thin shell show that the inverse linear divergences
in the local energy density which occur in either
order $\lambda$ or $\lambda^2$ cancel between inside and outside, when one
computes the total energy,
while the divergence encountered at $n=3$:
\be u^{(3)}\sim \frac{\lambda^3}{a^7}\frac1{144\pi^2}\Gamma(0),
\ee
is correctly given by
\be
u^{(3)}\sim -\frac{\lambda^3}{144\pi^2 a^7}\ln\frac{r-a}a,
\ee
as shown by explicit calculation, because
\be
\sum_{l=0}^\infty \frac1\nu e^{-\nu\chi}\sim-\ln\chi,\quad\chi\to0.
\ee
The integral of $u^{(3)}$, however, is finite, so this does not
signal any difficulty with the total energy in order $\lambda^3$.

\subsection{Surface energy}
However, as discussed first by Dowker, Kennedy and Critchley
\cite{Dowker:1978md,Kennedy:1979ar}, and later elaborated by
Saharian and Romeo \cite{saharian},
and put in a broader context by Fulling \cite{fulling}, for situations
when other than Neumann or Dirichlet boundary conditions apply, an
additional term must be supplied in calculating the energy, a term
 which resides entirely on the surface.  For the case of the general
stress tensor (\ref{stresstensor}), that extra term is \cite{Milton:2004vy}
\be
\mathfrak{E}=-\frac{1-4\xi}{2i}\int_S d\mathbf{S}\cdot \bm{\nabla}G(x,x')
\bigg|_{x'\to x},
\label{st}
\ee
where the direction of the normal is out of the region in question,
which arises from the $T^{0i}$ component of the stress tensor,
and from $\partial_\mu T^{\mu\nu}=0$.
The total energy in a given region $V$ bounded by a surface $S$
is not, therefore, just the integral of the local energy density, but has 
this extra contribution \cite{Milton:2004vy}:
\be
\mathcal{E}=
\int_V (d\mathbf{r})\langle T^{00}\rangle+\mathfrak{E}=\frac1{2i}\int_V(d\mathbf{r})
\int\frac{d\omega}{2\pi}2\omega^2\mathcal{G}(\mathbf{r,r}) e^{-i\omega\tau},
\label{totenergy}
\ee
which is independent of $\xi$.
The latter expression has a rather evident interpretation in terms
of summing zero-point energies.  (For example, see Appendix A
of Ref.~\cite{miltonbook}.) As there, we have inserted a temporal 
point-splitting
regulator $\tau$ as first introduced in Ref.~\cite{Milton:1978sf}.
The limit $\tau\to0$ must be taken at the end of the calculation.
Of course, the total energy in all space is just the sum of the integrated
local energy densities in each region, because the sum of the inside and
outside surface energy for each surface is zero, owing to the continuity
of the Green's function and its normal derivative across each surface for the
nonsingular potential (\ref{smoothpot}).  This is not the case for the singular
potential (\ref{deltapot}).

\subsection{Shell energy}
In the limit of $h\to\infty$ for the region in the shell, $a_-<r,r'<a_+$,
Eq.~(\ref{2.10}) becomes [$\lambda h\gg (\kappa a)^2$, which is the
opposite limit to that taken in Sec.~\ref{sec:int}]
\be
g_l\to \frac1{2\kappa rr'}\frac{e_l(\kappa a)s_l(\kappa a)}{1+\frac\lambda
{\kappa a^2}e_l(\kappa a)s_l(\kappa a)}
\bigg[\cosh\frac{\sqrt{\lambda h}}a(r-r')
+\cosh\frac{\sqrt{\lambda h}}a(r+r'-a_+-a_-)\bigg].
\ee
(Recall we are disregarding the physically spurious planar surface
divergences, discussed in Sec.~\ref{sec:int}, which in any case disappear
for the conformal value $\xi=1/6$.)
In the thin shell limit this leads to an energy density in the
shell nearly independent of $r$ ($\delta\to0$), leading to the energy
($y=|x|$, $\epsilon=\tau/a$, $\tau$ now being the Euclidean time)
\bea
E_s=\int(d\mathbf{r}) \langle T^{00}\rangle
=\frac\lambda{2\pi a^2}(1-4\xi)\sum_{l=0}^\infty (2l+1)
\frac12\int_{-\infty}^\infty dx
\frac{I_\nu(y)K_\nu(y)}{1+\frac{\lambda}{a} I_\nu(y)K_\nu(y)}
e^{ix\epsilon}.\label{annen}
\eea
However, we have to include the surface term in the shell, in the thin
shell limit, which from Eq.~(\ref{st}) is
\bea
\mathfrak{E}_{s}
=-\frac{1-4\xi}{2i}\int d\mathbf{S}\cdot \bm{\nabla} G(\mathbf{r},t;
\mathbf{r'},t-\tau)\bigg|_{\mathbf{r'=r},r=a_-}^{\mathbf{r'=r},r=a_+},
\eea
where the change in sign in the normal is now encompassed in the limits.
$\mathfrak{E}_s$ exactly cancels $E_s$ in the thin shell limit: 
$\mathcal{E}_s=E_s+\mathfrak{E}_{s}\to0$, 
because the total energy of the shell is given by Eq.~(\ref{totenergy})
\bea
\mathcal{E}_{s}=\frac1{2i}\int_{\text{shell}}(d\mathbf{r})
\int\frac{d\omega}{2\pi}2\omega^2 \mathcal{G}(\mathbf{r,r})
e^{-i\omega\tau},
\eea
which clearly vanishes as the thickness of the shell $\delta\to0$,
because the integral is proportional to the volume of the shell.
However, we shall see shortly that $E_s$ in Eq.~(\ref{annen}) 
 plays a special role. [The reader should note that if we set $\tau=0$ 
expressions such as Eq.~(\ref{annen}) fail to converge.  
As we shall see shortly, if this point coincident limit  is expanded in powers
of $\lambda$, divergences in the total energy
 occur in orders $\lambda$ and $\lambda^3$.]

\subsection{Total energy of $\lambda$ sphere}
Likewise, if one integrates the interior and exterior energy density
in the thin shell limit, one gets,
for arbitrary $\xi$,
\be
E_{\rm in}+E_{\rm out}=\mathcal{E}-\mathfrak{E},\label{intenergy}
\ee
where the total energy is that found in 
Ref.~\cite{Bordag:1998vs,Milton:2004vy}, ($y=|x|$)
\be
\mathcal{E}=-\frac1{2\pi a}\sum_{l=0}^\infty(2l+1)\frac12\int_{-\infty}^\infty
 dx\,y
\frac{d}{dy}\ln\left[1+\frac{\lambda}{a} I_\nu(y)K_\nu(y) \right]
e^{i x\epsilon},
\label{totale}
\ee
and where the extra term $\mathfrak{E}$
is precisely the sum of the surface terms for the inside and outside regions.
Again, Eq.~(\ref{intenergy}) is a restatement of Eq.~(\ref{totenergy}), and
 $\mathcal{E}$ is exactly that obtained from the integral of the 
Green's function, as in Eq.~(\ref{totenergy}).

The above can be verified by computing the pressure.
This is obtained by evaluating the radial-radial component of the stress tensor
(\ref{stresstensor}), or
\be
T_{rr}=\frac12\left[(\partial_r\phi)^2+(\partial_0\phi)^2-(\bm{\nabla}_\perp
\phi)^2\right]-\xi(\partial_0^2-\nabla_\perp^2)\phi^2,
\ee
or more precisely, the discontinuity of the vacuum expectation value of this
across the surface:
\bea P&=&\langle T_{rr}\rangle_{\mathrm{in}}-\langle T_{rr}\rangle_{\mathrm{out}}
\nonumber\\
&=&\frac12\int_{-\infty}^\infty \frac{d\zeta}{2\pi}
\sum_{l=0}^\infty \frac{2l+1}{4\pi}
\left[\partial_r\partial_{r'}-\kappa^2-\frac{l(l+1)}{r^2}\right]g(r,r')
\bigg|_{r'=r=a+}^{r'=r=a-} e^{i\zeta\tau},
\eea
with $\kappa =|\zeta|$.
We note that the $\xi$ dependence drops out immediately, and the terms not
involving radial derivatives of Bessel functions cancel between inside and 
outside, leaving us with, in the thin-shell limit,
\be
P=-\frac\lambda{8\pi^2 a^5}
\frac12\int_{-\infty}^\infty dx\sum_{l=0}^\infty (2l+1)
\frac{y[I_\nu(y)K_\nu(y)]'-I_\nu(y)K_\nu(y)}{1
+\frac\lambda{a}I_\nu(y)K_\nu(y)}e^{ix\epsilon},
\ee
which, 
when multiplied by the area of the sphere to give the stress on the surface,
is exactly that obtained by differentiating the total energy (\ref{totale})
with respect
to the radius (with $\lambda$ fixed), 
\be
-\frac{\partial\mathcal{E}}{\partial a}=4\pi a^2 P.
\ee
(The scaling law for $\lambda$ was incorrectly stated in 
Refs.~\cite{Milton:2004ya,Milton:2004vy}, as was the sign of the pressure.)

However, there is more to say here.  As noted above, the integral of the
local energy, inside and outside the sphere, is finite perturbatively,
because of cancellations between inside and outside, for the conformally
coupled scalar.
But it is well known that divergences occur in the
total energy at order $\lambda^3$ 
\cite{Bordag:1998vs,Graham:2003ib,Milton:2004ya}.  
These evidently must arise from the
surface term (\ref{st}).  So let us consider the latter, which
 is given in  the outside region by
\be
\mathfrak{E}=a^2(1-4\xi)\sum_{l=0}^\infty 2\nu\frac12\int_{-\infty}^\infty 
\frac{d\zeta}{2\pi}\frac{\partial}
{\partial r}g_l(r,r')\bigg|_{r=r'=a+}e^{i\zeta\tau}.\label{surfout}
\ee

In the strong coupling limit, there is, of course, no surface term.
This is because then
\be
r,r'>a:\quad g_l(r,r')=\frac1{\kappa rr'}\left[s_l(\kappa r_<)e_l(\kappa r_>)
-\frac{s_l(\kappa a)}{e_l(\kappa a)}e_l(\kappa r)e_l(\kappa r')\right],
\ee
which vanishes on the surface, and has a derivative proportional to the
Wronskian.

In general, in the thin-shell limit, the sum of the inside and outside surface 
terms is given by
\be
\mathfrak{E}=\frac\lambda{2\pi a^2}(1-4\xi)\frac12
\int_{-\infty}^\infty dx \sum_{l=0}^\infty (2l+1)
\frac{I_\nu(y)K_\nu(y)}{1+\frac\lambda{a}I_\nu(y)K_\nu(y)}e^{i x\epsilon},
\ee
because the free term in the Green's function cancels between inside and out.
It is apparent that the strong coupling limit of this vanishes, being just
a delta function.
Perhaps not remarkably, this is precisely the same as the integrated local
shell energy $E_s$, (\ref{annen}).  
Thus the surface energies within and outside
the shell regions cancel. (As noted above, this is a general result, 
because the Green's 
function and its normal derivative are continuous across each surface at $a_+$,
$a_-$.  We have explicitly verified this using (\ref{goutside}), 
(\ref{2.10}), and the Wronskian for the Bessel functions.)
For weak coupling, we expand this in powers of $\lambda$.
Again, the easiest way to isolate the asymptotic behavior is to use
the uniform asymptotic expansion,
\be
\nu\to\infty:\quad I_\nu(x)K_\nu(x)\sim \frac{t}{2\nu}.
\ee
This yields the following expression for the $n$th term in the total
surface energy, if we set $\epsilon=0$ and regard $n$ as a continuous
variable, analytically continuing from $\mbox{Re}\, n>3$,
\be
\mathfrak{E}^{(n)}\sim -\frac{(-1)^n}{2\sqrt{\pi}a}(1-4\xi)
\left(\frac{\lambda }{2a}\right)^n
\frac{\Gamma\left(\frac{n-1}2\right)}{\Gamma\left(\frac{n}2\right)}(2^{n-2}-1)
\zeta(n-2).
\ee
 Note that
this expression vanishes for $n=2$; we must keep subleading corrections to see
the order $\lambda^2$ term in the energy
arising from the surface energy.  However, for $n=3$ we obtain
for the conformal value, $\xi=1/6$,
\be
\mathfrak{E}^{(3)}\sim \frac{\lambda^3}{24\pi a^4}\zeta(1),
\ee
precisely the divergent term in the energy given in Ref.~\cite{Milton:2004vy},
first found by the heat kernel calculation of Bordag, Kirsten, and 
Vassilevich \cite{Bordag:1998vs}.
Alternatively if we keep the point-split regulator, we find the
$n=2$ term still vanishes,
\be
\mathfrak{E}^{(2)}\sim -\frac{\lambda^2}{24\epsilon a^3}\frac{i}\epsilon
\int_{-\infty}^\infty \frac{dz}z\frac1{z^2+1}=0,\ee
which we argue is imaginary and the integrand odd, 
if we interpret the integral as its
principal value.  The same phenomenon resulted in the finiteness of the
energy for the strong coupling limit---the Dirichlet sphere---see 
Ref.~\cite{Milton:2002vm,Milton:2004ya}. 
The divergence at $n=3$ persists:
\be
\mathfrak{E}^{(3)}\sim\frac{\lambda^3}{12\pi a^4}\ln\epsilon.
\ee
Thus, as shown in Ref.~\cite{Milton:2004vy}, the total energy term
$\mathcal{E}^{(2)}$ from Eq.~(\ref{totale}),
by expanding the logarithm to order $\lambda^2$ and applying the
derivative operator,  is unambiguously finite:
\be
\mathcal{E}^{(2)}=\frac{\lambda^2}{32\pi a^3}.
\ee 
On the other hand, $\mathcal{E}^{(3)}$ is unambiguously divergent.
\subsection{Other values of $\xi$}
We have shown, as expected, that for the global energy any value of
$\xi$ can be used.  Fulling \cite{fulling} has suggested that a value
different from the conformal value may have advantages.  Thus, because
the surface (or shell) energy is proportional to $1-4\xi$, 
it vanishes for $\xi=1/4$,
as do most of the terms in the energy density (\ref{uout}).
In that case, however, the surface divergences in the local energy density
are intensified, which does not seem to
account for the divergence in the total
energy in order $\lambda^3$
[because the inverse linear divergence seen in Eq.~(\ref{surfdivn}) 
would cancel between the inside and the outside], 
and worse leads to a divergence in $\lambda^2$.  It is probably
not surprising that the situation is most satisfactory only in the conformally
coupled case, because it is well known that any other choice leads to
more singular quantum corrections \cite{Milton:1972cc}.

\section{Conclusions}
We have derived the Green's function in general for a  massless scalar field
for a spherically symmetric step-function potential.  By taking the
limit as the width of the step function goes to zero, we recover a 
$\delta$-function shell potential, for which we consider both the weak and
strong coupling limits.  The latter corresponds to a Dirichlet shell, for
which we recover the well-known results for the energy density and total
energy.  For weak coupling, we derive for the first time the behavior of
the energy density as the surface is approached.  We also examine the
energy content of the shell itself, both for a thick shell, and in the thin
shell limit.  In the former case there are unphysical nonconformal
divergences in the local energy density as the surfaces are approached
which are exactly those found for
a plane slab.  These are most easily eliminated by using the conformal
stress tensor.  The inner and outer surfaces of a thick
shell contribute a vanishing net surface energy, but there is a net effective 
surface energy in the thin shell limit, to be added to the integrated
local energy density for the inside and outside regions, 
which is exactly the integrated local energy density
of the shell.  

The situation is summarized as follows.  The integrated local energy density
inside, outside, and within the shell is $E_{\rm in}$, $E_{\rm out}$, and
$E_s$, respectively.  The total energy of a given region is the sum of
the integrated local energy and the surface energy bounding that region:
\begin{subequations}
\bea
\mathcal{E}_{\rm in}&=&E_{\rm in}+\mathfrak{E}_-,\\
\mathcal{E}_{\rm out}&=&E_{\rm out}+\mathfrak{E}_+,\\
\mathcal{E}_s&=&E_s+\mathfrak{E}'_++\mathfrak{E}'_-,
\eea
\end{subequations}
where $\mathfrak{E}_\pm$ is the outside (inside) surface energy on the
surface at $r=a_\pm$, while $\mathfrak{E}_\pm'$ is the inside (outside) surface
energy on the same surfaces.  Because for a nonsingular potential the surface
energies cancel across each boundary,
\be
\mathfrak{E}_++\mathfrak{E}_+'=0,\quad\mathfrak{E}_-+\mathfrak{E}_-'=0,
\ee
the total energy is 
\be
\mathcal{E}=\mathcal{E}_{\rm in}+\mathcal{E}_{\rm out}+\mathcal{E}_s
=E_{\rm in}+E_{\rm out}+E_s.
\ee
In the singular thin shell limit, the integrated local shell energy
is the total surface energy of a thin Dirichlet shell:
\be
E_s=\mathfrak{E}_++\mathfrak{E}_-\ne 0.
\ee
This shell energy, for the conformally coupled
theory, is finite in second order in the
coupling (in at least two plausible regularization schemes), 
but diverges in third order.  We show that the latter precisely
corresponds to the known divergence of the total energy in this order.
Thus we have established the suspected correspondence between surface
divergences and divergences in the total energy, which has nothing to
do with divergences in the local energy density as the surface is approached.
This precise correspondence should enable us to absorb such 
global divergences in
a renormalization of the surface energy, and should lead to further advances
of our understanding of quantum vacuum effects.

\begin{acknowledgments}
We are grateful to
the US Department of Energy for partial financial support of this research.
We thank Stuart Dowker,
Steve Fulling, Noah Graham,  Klaus Kirsten,  and Prachi Parashar
 for helpful conversations.
\end{acknowledgments}


\begin{thebibliography}{99}

\bibitem{Brown:1969na}
  L.~S.~Brown and G.~J.~Maclay,
  Phys.\ Rev.\  {\bf 184}, 1272 (1969).


\bibitem{Casimir:1948dh}
  H.~B.~G.~Casimir,
  Kon.\ Ned.\ Akad.\ Wetensch.\ Proc.\  {\bf 51}, 793 (1948).

\bibitem{Milton:2002vm}
  K.~A.~Milton,
  Phys.\ Rev.\ D {\bf 68}, 065020 (2003)
  [arXiv:hep-th/0210081].

\bibitem{Actor:1996zj}
  A.~A.~Actor and I.~Bender,
  Fortsch.\ Phys.\  {\bf 44}, 281 (1996).


\bibitem{Deutsch:1978sc}
  D.~Deutsch and P.~Candelas,
  Phys.\ Rev.\ D {\bf 20}, 3063 (1979).


\bibitem{Boyer:1968uf}
  T.~H.~Boyer,
  Phys.\ Rev.\  {\bf 174}, 1764 (1968).



\bibitem{Bender:1994zr}
  C.~M.~Bender and K.~A.~Milton,
  Phys.\ Rev.\ D {\bf 50}, 6547 (1994)
  [arXiv:hep-th/9406048].

\bibitem{graf}
F. Bernasconi, G. M. Graf, and D. Hasler, Ann. Inst. H. Poincar\'e \textbf{4},
1001 (2003) [arXiv:math-ph/0302035].


\bibitem{Graham:2003ib}
  N.~Graham, R.~L.~Jaffe, V.~Khemani, M.~Quandt, O.~Schroeder and H.~Weigel,
  Nucl.\ Phys.\ B {\bf 677}, 379 (2004)
  [arXiv:hep-th/0309130], and references therein.


\bibitem{Milton:2004ya}
  K.~A.~Milton,
  J.\ Phys.\ A {\bf 37}, R209 (2004)
  [arXiv:hep-th/0406024].


\bibitem{Milton:2004vy}
  K.~A.~Milton,
  J.\ Phys.\ A {\bf 37}, 6391 (2004)
  [arXiv:hep-th/0401090].


\bibitem{Bordag:1998vs}
  M.~Bordag, K.~Kirsten and D.~Vassilevich,
  Phys.\ Rev.\ D {\bf 59}, 085011 (1999)
  [arXiv:hep-th/9811015].
  

\bibitem{Bordag:2004rx}
  M.~Bordag and D.~V.~Vassilevich,
  Phys.\ Rev.\ D {\bf 70}, 045003 (2004)
  [arXiv:hep-th/0404069].

\bibitem{Scandurra:1998xa}
  M.~Scandurra,
  J.\ Phys.\ A {\bf 32}, 5679 (1999)
  [arXiv:hep-th/9811164].


\bibitem{Mckenzie-Smith:2004hg}
  J.~J.~McKenzie-Smith and W.~Naylor,
  Phys.\ Lett.\ B {\bf 610}, 159 (2005)
  [arXiv:hep-th/0412156].



\bibitem{barton} G. Barton, J. Phys. A {\bf 37}, 1011 (2004); J. Phys. A
{\bf 34}, 4083 (2001y).

\bibitem{Graham:2002yr}
N. Graham and K. D. Olum,
Phys. Rev. D {\bf67}, 085014 (2003)
[arXiv:hep-th/0211244] [Erratum {\bf 69}, 109901 (2004)].

\bibitem{Milton:1979yx}
  K.~A.~Milton,
  Ann.\ Phys.\ (N.Y.)  {\bf 127}, 49 (1980).


\bibitem{Schwartz-Perlov:2005ds}
  D.~Schwartz-Perlov and K.~D.~Olum,
Phys.\ Rev.\ D {\bf 72}, 065013 (2005)
  [arXiv:hep-th/0507013].


\bibitem{Dowker:1978md}
  J.~S.~Dowker and G.~Kennedy,
  J.\ Phys.\ A {\bf 11}, 895 (1978).

\bibitem{Kennedy:1979ar}
  G.~Kennedy, R.~Critchley and J.~S.~Dowker,
  Ann.\ Phys.\ (N.Y.)  {\bf 125}, 346 (1980).

\bibitem{saharian} A. Romeo and A. A. Saharian, J. Phys. A {\bf 35}, 1297 
(2002) [arXiv:hep-th/0007242]; Phys. Rev. D {\bf 63}, 105019 (2001)
[arXiv:hep-th/0101155].

\bibitem{fulling} S. A. Fulling, J. Phys. A {\bf 36}, 6529 (2003)
[arXiv:quant-ph/0302117].

\bibitem{miltonbook}
K. A. Milton, {\it The Casimir Effect: Physical Manifestations of Zero-Point
Energy} (World Scientific, Singapore, 2001).

\bibitem{Milton:1978sf}
  K.~A.~Milton, L.~L.~DeRaad, Jr., and J.~S.~Schwinger,
  Ann.\ Phys.\ (N.Y.)  {\bf 115}, 388 (1978).

\bibitem{Milton:1972cc}
  K.~A.~Milton,
  Phys.\ Rev.\ D {\bf 4}, 3579 (1971).

\end{thebibliography}
\end{document}